# Self-Excitation: An Enabler for Online Thermal Estimation and Model Predictive Control of Buildings

Peter Radecki and Brandon Hencey

*Abstract*—This paper investigates a method to improve buildings' thermal predictive control performance via online identification and excitation (active learning process) that minimally disrupts normal operations. In previous studies we have demonstrated scalable methods to acquire multi-zone thermal models of passive buildings using a gray-box approach that leverages building topology and measurement data. Here we extend the method to multi-zone actively controlled buildings and examine how to improve the thermal model estimation by using the controller to excite unknown portions of the building's dynamics. Comparing against a baseline thermostat controller, we demonstrate the utility of both the initially acquired and improved thermal models within a Model Predictive Control (MPC) framework, which anticipates weather uncertainty and time-varying temperature set-points. A simulation study demonstrates self-excitation improves model estimation, which corresponds to improved MPC energy savings and occupant comfort. By coupling building topology, estimation, and control routines into a single online framework, we have demonstrated the potential for low-cost scalable methods to actively learn and control buildings to ensure occupant comfort and minimize energy usage, all while using the existing building's HVAC sensors and hardware.

## I. INTRODUCTION

Buildings' heating, ventilating, and air conditioning (HVAC) systems represent a large opportunity for energy savings in the US and abroad [1]. To date few scalable and cost-effective methods exist for optimally controlling buildings because of the diversity in their construction, usage, and heating/cooling equipment [2]. As demonstrated in our previous study [3], gray-box methods, utilizing building topology plus online temperature and weather data, can be used as a low-cost, scalable way to learn passive thermal models from measured data. Unfortunately the data typically collected from buildings often does not contain enough information content to calibrate and learn a control-oriented model with useful predictive capabilities [4], [5], [6], [7], [8]. Based on our previous work [9], by adding self-excitation via HVAC actuators, we propose a managed online framework for model estimation, active learning, and predictive control that is scalable to a wide range of buildings.

Due to recent trends in both new and retrofit construction, it is common to see high efficiency components, distributed sensing, networked devices, and computational power in buildings. Typically absent are model predictive controllers from the Building Automation System (BAS) [10]. Commissioning is often done heuristically and the effectiveness of the control architecture is typically a function of the software interface and the technician's understanding of the building. The level of insight, complexity, and tuning required to deliver a high performance BAS is labor and expertise intensive; therefore, heuristic and intuition-driven approaches are not scalable or economical for most buildings [11]. In an ideal BAS, a predictive controller would understand thermal dynamics, occupant usage, weather predictions, and typical disturbances in order to save energy while improving occupant comfort. As has been investigated by numerous researchers [12], [13], [14] Model Predictive Control (MPC) provides a practical method to minimize energy while ensuring occupant comfort in the presence of changing weather and disturbance uncertainties. MPC has been shown to be especially effective in dealing with smart grid driven time-of-usage costs and Demand Response (DR) in buildings [14], [15].

The main difficulty in implementing MPC for HVAC in buildings is acquiring an accurate model of both the thermal dynamics and disturbances for the building [2], [16]. For real-time control the model should be as simple as possible, however modeling discrepancies can cause MPC performance degradation. Gunay et. al. showed that poor disturbance or occupancy models can cause MPC to perform worse than a typical thermostat [13]. Behl et.al. explored the sensitivity of temperature predictions and controller energy consumption to perturbations of model parameters [17]. Massoumy et al. compared MPC to rule-based (thermostat) control for various accuracy models and found MPC would only outperform rule-based control given an accurate model [18].

In [3] we proposed a multi-mode Unscented Kalman Filter (UKF) that learned buildings' multi-zone thermal dynamics and detected unknown time varying thermal loads. Given the multi-zone structure of the building, the UKF initially estimated parameters of that structure using low disturbance periods, and then was augmented to track

This work was supported by the Department of Defense (DoD) through the National Defense Science & Engineering Graduate Fellowship (NDSEG) Program.

P. Radecki and B. Hencey were with the Sibley School of Mechanical and Aerospace Engineering, Cornell University, Ithaca, NY, 14850 USA. Currently Radecki is a Transmission Development Engineer at General Motors Powertrain in Detroit, Michigan, *email:* ppr27@cornell.edu. Currently Hencey is with the Air Force Research Laboratory in Dayton, Ohio, *email:* bhencey@gmail.com.

thermal disturbances while continuing to refine its thermal parameter estimates. Models learned on data with high information content had good predictive capability. As proposed, an online BAS framework is susceptible to failure in several ways: correct operation of the MPC is completely dependent on the learned thermal model, and the thermal model estimation process is dependent on both the quality of excitation in measured data and the utility of the underlying model framework.

Learning the dynamics of any system from measured data requires excitation of the given system; without sufficient excitation, the system parameters are unobservable. In an operational building, induced excitation is often necessary—due to poor zone segregation, bad signal to noise ratio, or poor data observability—but difficult to manage without wasting energy or annoying occupants [4].

Thermal excitation of buildings is an area of recent interest. Cigler used subspace methods for learning numerical models of buildings and discussed the importance of excitation in data while learning models [19]. Bengea looked at closed loop sensitivity of learned parameters to control performance and noted the importance of persistent excitation in the measured data [5].

Various excitation schemes have been examined to improve the information content of measured data. Agbi looked at the identifiability of parameters by designing experiments such as square wave and sinusoidal experiments and derived information criteria in measured data [20]. Gorni used bounded pseudo-random inputs for an estimation procedure to improve initial parameter estimates [21]. Many other researchers have demonstrated how, given a high fidelity Energy Plus model, a system identification routine in Matlab can be run using MLE+ to excite the system with pseudo-random binary inputs to generate control-oriented models [22]. Aswani split the learning into parametric (resistor capacitor model) and non-parametric (heating disturbances) components inside Learning Based MPC. Insufficient excitation was compensated with a Bayesian prior for parameter estimates [7].

Data-driven modeling is essential for MPC but a difficult challenge for buildings. Privara highlighted three main challenges in data-driven modeling for MPC: 1) data violates typical persistent excitation requirements, 2) increased model complexity increases length of learning time and suitable experiments may be expensive, 3) measured temperature signals are often co-linear [4]. Li, who did the most recent extensive literature survey on building modeling for control and operation, concluded that generation of appropriate excitation signals while considering the characteristics and constraints of building energy systems is an urgent research topic [6]. Studies to date have demonstrated the importance of self-excitation and shown the benefits of a learned model coupled to MPC. However, a comprehensive framework that learns over time and automatically identifies appropriate self-excitation has yet to be presented.

Thermal excitation of a building consumes energy, may disrupt occupants, and has the potential to damage or accelerate wear on HVAC systems. We propose a framework for online self-excitation of the HVAC system that provides the information necessary to learn and update building models while not wasting energy, unnecessarily disrupting occupants, or breaking equipment. By building upon our initial study presented in [9] the main contributions of this paper include:

- Development of an Experiment Generator that determines, based on the current model, which zones should be excited;
- Development of an Experiment Selector that automatically selects and runs experiments while obeying energy and occupant constraints;
- Observability analysis confirming and explaining the basis for our excitation approach.

The Experiment Generator and Selector are wrapped into an overall on-line framework that includes an UKF and MPC for estimation and control.

We propose a framework to demonstrate the utility of coupling gray-box estimation methods with predictive controllers to create an online deployable BAS. After a brief explanation of the chosen underlying thermal model and Kalman Filter parameter estimate representation, we highlight the Model Predictive Controller with soft-constraint temperature bounds, excitation generation, and estimation monitoring. A 2-zone building simulation is developed to demonstrate the entire framework actively learning, improving occupant comfort, and reducing energy use compared to a baseline thermostat controller. An observability analysis sheds light on what mathematically occurs during self-excitation of the building. We conclude by discussing limitations of our framework and directions for future research.

II. THERMAL MODEL ESTIMATION

The thermal model, estimator, and controller shown below match that presented in our early investigation [9] and are included here, partially in verbatim, as a basis upon which to derive the new excitation contributions of this paper, and such that the entire online framework may be coherently seen and understood.

*A. RC Thermal Model Parameterization*

A resistor-capacitor (RC) network is used to model the conduction, convection, and mass transfer thermal dynamics occurring in the building. Radiation is assumed to be linearly approximated in the RC model. An additive heat coefficient term is included to model the influence of the heating system in the building. Although our explanation is brief; a more thorough treatment is offered in [3] or [23].

In order to model buildings with an RC network the building must be split into zones with temperature $T_i$ based on the building's structure—for this study it is assumed that each zone has independent heating control. The heat flux added to zone $i$ is given as the sum of contributions from adjacent zone(s) $j$ plus a contribution from zone heater $b_i$, where $R_{ij}$ is the insulation or thermal resistance between zones. The temperature rate of change $\dot{T}_i$ is the heat flux divided by is the thermal mass of the zone $C_i$. The rate of change of temperature of zone $i$ due to connection with zone(s) $j$ is

$$\dot{T}_i = \sum_j (T_j - T_i)/(R_{ij} C_i) + b_i/C_i. \quad (1)$$

This formulation can be expanded to build a full linear state space representation of the thermal dynamics (2) where each thermal zone is a node in the RC network. Matrix $A$ represents the heat exchange between zones, $\bar{T}(t)$ is a vector of node temperatures, $B$ represents the total heat output of the heater per zone, and $\bar{u}(t)$ is the control vector describing the fraction of the heater's output varying between 0 and 1. The formulation of matrix $A$ is included because its structure is used later in the self-excitation explanation. For a general thermal network with $n$ nodes, the $A$ matrix can be constructed in (3) as a simple undirected weighted graph with: nodes $N \coloneqq \{1,2,\ldots,n\}$ that are assigned capacitances $C_i$ and temperatures $T_i$; edges $E \subset N \times N$ that connect adjacent nodes; and weights $\{R_{ij} \forall (i,j) \in E : R_{ij} = R_{ji}\}$ that are assigned resistances. Lastly, the diagonal matrix $B$ is defined per node by the maximum heater output $b_i$ divided by the thermal capacitance $C_i$ as shown in (4). External temperature is modeled as a node in the thermal network with infinite capacitance whose temperature is modified by a weather forcing function.

$$\dot{\bar{T}}(t) = A\bar{T}(t) + B\bar{u}(t) \quad (2)$$

$$A = \begin{cases} A_{ij} = 0 & \text{if } i \neq j, (i,j) \notin E \\ A_{ij} = \dfrac{1}{C_i R_{ij}} & \text{if } i \neq j, (i,j) \in E \\ A_{ij} = -\sum_{l \neq i} A_{il} & \text{if } i = j \end{cases} \quad (3)$$

$$B = \text{diag}_{i=1\ldots n}\{b_i/C_i\} \quad (4)$$

In order to provide stable estimation and enforce the underlying dynamics, a minimal parameterization of the RC model must be determined and used by the parameter estimation algorithm. For this we leverage the minimal parameterization described in [3]. Any graphs containing cycles (non-tree graphs) are analyzed to multiplicatively cancel out redundant parameters. Each unique RC product is estimated as a single parameter $p_k$, and each controllable zone's heat coefficient $b_i/C_i$ is estimated as parameter $q_l$. For notational purposes the final set of parameters are grouped into two vectors $\bar{p}$ and $\bar{q}$.

### B. UKF Estimation Procedure

Merging data measured from the building, knowledge about disturbances, and first-principles parameterization of the building's topology inside an online estimation algorithm provides us with a scalable framework (Fig. 1) that can be deployed on buildings. An Unscented Kalman Filter provides a numerically stable and near-optimal[1] way to estimate parameters, disturbances, and temperature states.

Temperature states are included in the estimation routine because temperature measurements in buildings are not perfect: sensors may be slightly miscalibrated, an air volume may not be thoroughly mixed, and individual thermal zones may have some unobservable temperature gradient. Parameter estimation is achieved in the Kalman Filter by augmenting the parameters onto the state vector of node temperatures $\bar{T}$ as

$$\hat{x} = \begin{bmatrix} \bar{T} \\ \bar{p} \\ \bar{q} \end{bmatrix}.$$

This augmentation causes the state update function $Ax(t)$ to become a non-linear relationship—parameters $p_k$ and $q_l$ are being multiplied by states $T_i$ and inputs $u_i$, respectively. This non-linearity, written as $x(k+1) = f(x(k), u(k))$, is what necessitates the Unscented Kalman Filter prediction equations. The measurement function is linear because each temperature is directly measured, so the regular linear Kalman Filter equations are used for the measurement update step. For thorough UKF derivation and explanation, please see [24] and [25].

### III. CONTROLLER ARCHITECTURE

#### A. Baseline Thermostat Control

To evaluate the performance of model-based control a baseline thermostat controller, similar to what is widely deployed in most current buildings, was developed for the simulation. A winter office building heating scenario was chosen for the simulation study. Minimum and maximum temperature set points were chosen for occupied and unoccupied building zones and tied to a schedule which the controllers could use.

Additionally a turn-on preheat time schedule was added to the temperature profile for the thermostat controller. This preheat buffer time was sized to match the time the heater would take to heat the building from its unoccupied temperature set point to its occupied temperature set point if the outside temperature was 32 degrees. Additionally a hysteresis timer of 15 minutes between turn-on/turn-off events was included to prevent the controller from chattering on and off at every integration time step. Alternatively a Proportional-Integral controller could have been used, but would have required the same preheat buffer and performed similarly to the thermostat for our simple 2-zone building.

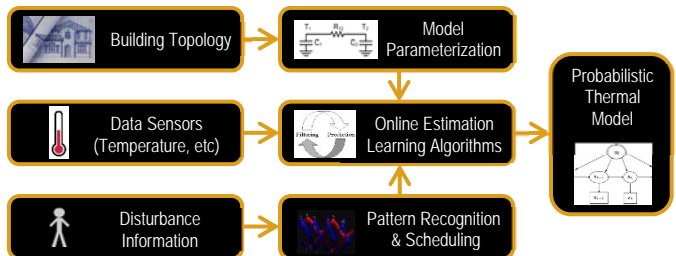

Fig. 1. Thermal Model Estimation Overview

---

[1] UKF is optimal for linear systems but sub-optimal for non-linear systems.

### B. Model Predictive Controller

A model-based control method is presented to demonstrate the efficacy of an improved system thermal model. Model Predictive Controller computes an optimal control input to minimize some cost function over a specified horizon given state and dynamics constraints. The formulations given here follow standard convention as presented in [26]. The MPC developed will minimize energy use, saving money, and bound occupant discomfort by ensuring temperatures stay within comfort bounds. The following section starts with a simple quadratic cost and then adds in time-varying temperature bounds for occupant comfort. Energy cost is assumed uniform; time of use charges could readily be added for industrial users.

Dynamics constraints for this problem are based on the thermal model estimated by the UKF. External weather, given at time step intervals by the most recently available prediction, is considered a constrained state and thus included as an additive term $T_{ext}$ in order to minimize the state dimension. An execution time step interval of 15 minutes was selected. The discrete state space system is

$$T(k+1) = AT(k) + B_{ext}T_{ext}(k) + B_{ctrl}u(k) \quad (5)$$

where $T(k)$ is the temperature vector, $u(k)$ is the control input, $A$ and $B_{ext}$ are populated with RC parameters $p_k$, and $B_{ctrl}$ is populated with heat coefficients $q_l$ that are learned by the UKF. All control inputs $u_i(k)$ are subject to the constraint set $0 \le u_i(k) \le 1$, and $T(0)$ is set to the current temperature.

Equation (6) gives an example cost function $J$ to be minimized over horizon $h$ given $n$ controllable temperature states. It is composed of a reference quadratic (2-norm) tracking term multiplied by weight $Q$ plus a 1-norm control effort term multiplied by weight $R$.

$$J = Q\sqrt{\frac{1}{nh}\sum_{k=1}^{h}(T(k)-r(k))^T(T(k)-r(k))} \\ + R\sum_{k=0}^{h-1}|u(k)| \quad (6)$$

Utilizing the given constraints and cost function, the MPC

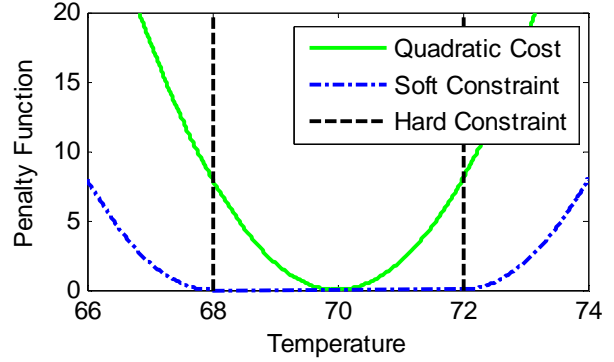

Fig. 2. Comparison of temperature cost functions.

problem can be posed as a disciplined convex optimization problem and solved in Matlab using CVX [27], [28]. Experimentation showed the Gurobi solver [29], available as third-party solver supported by CVX, to be 2 to 10 times faster at solving the MPC problem than CVX's pre-configured SeDuMi or SDPT3 solvers.

### C. Soft Constraints

The cost function given in (6) works well for pure reference tracking but poorly for typical energy saving climate control where one only desires that the temperature be kept between two bounds. References [26] and [30] present an alternative method utilizing soft-constraints. Hard constraints, such as keeping the temperature between upper and lower bounds, do not work well for dynamical system outputs because they can make the problem unsolvable under normal operational conditions. Consider if a thermal disturbance pushed the temperature far enough outside of the hard bounds that the temperature bounds could not be met at the next time step. In such a scenario the solver would fail to calculate an output. Simply speaking, the maximal or minimal control effort applied for one time step would fail to allow the system to satisfy the hard temperature constraints, and thus the controller would do nothing. By using soft constraints the controller is robust to this failure condition. If too far outside the bounds, the controller would simply apply maximal hard-constrained control effort until the temperature is within the soft-constraint bounds.

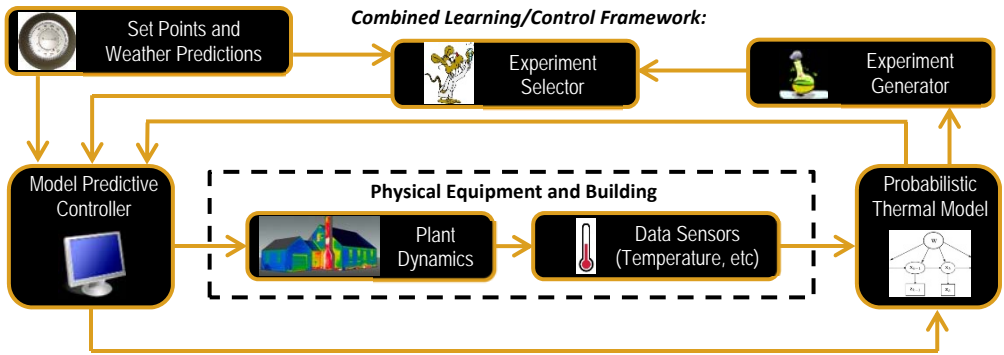

Fig. 3 Combined framework for parameter estimation, control, and self-excitation.

A comparison of quadratic (2-norm regulating 70°), soft-constraint (2-norm regulating between 68° and 72°) and hard-constraint (bounded between 68° and 72°) penalty functions are shown in Fig. 2. A naïve method of mathematically writing the soft-constraint might be given by $J_Q = \|\min(T(k) - T_{min}, 0)\|^2 + \|\max(T(k) - T_{max}, 0)\|^2$ and preclude the use of any convex solver due to the minimum and maximum functions. By introducing an $n \times h$ size vector of slack variables $w$, the soft-constraint can be fully formulated in a convex cost function as shown in (7). Note that $T(k)$ is vector of length $n$ so $T$, $w$, $r_{min}$, $r_{max}$, and $u$ are all size $n \times h$. An additional cost-to-go term was included in $J$ where $r(k) = (r_{min}(k) + r_{max}(k))/2$. Given the discrete nature of the problem coupled with control constraints, the cost-to-go term helps the temperature to tend toward the middle of the bounded range.

*Minimize J:*

$$J = Q \sqrt{\frac{1}{nh} \sum_{k=1}^{h} w(k)^T w(k)} + R \sum_{k=0}^{h-1} |u(k)|$$

$$+ Q_{to\,go} \sqrt{\frac{1}{n}(T(h) - r(h))^T (T(h) - r(h))}$$

*Subject to:* (7)

$$T(k+1) = \Phi_A T(k) + \Gamma_{B_{ext}} T_{ext}(k) + \Gamma_{B_{ctrl}} u(k)$$
$$0 \leq u_i(k) \leq 1$$
$$T(k) + w(k) \geq r_{min}$$
$$T(k) - w(k) \leq r_{max}$$
$$w \geq 0$$

*Given numerical values for vectors:*

$$T_{ext}, T(0), r_{max}, r_{min}, r$$

## IV. MATLAB SIMULATED TEST BUILDING

### A. 2-zone Model

Most buildings do not have uniform usage. Office buildings, schools, and stores are generally occupied during the work-day while apartments and houses are generally occupied on nights and weekends. This non-uniform usage in the presence of varying external temperatures is what allows predictive controllers to outperform thermostat controllers—given a weather prediction and the current state, they can better anticipate how the building should be controlled. Our heating control example reflects the typical usage of a heated 2-zone office building in winter that is occupied during the day and vacant at night. The 2-zone building presented here was originally developed in [9].

Fig. 4 shows a diagram of the RC network representing the building while Table 1 contains the numeric values used in the main MPC versus thermostat simulation study. Values picked are given without units but chosen to correlate with typical cooling and heating time constants for a building measured in degrees Fahrenheit. Notice how zones 1 and 2 are weakly connected: $R_{12}$ has a higher thermal resistance than either $R_{13}$ or $R_{23}$. This may seem odd, but in fact is a realistic common occurrence, especially for buildings that have additions. The zones might share one wall with each other while each having 3 externals walls plus a roof causing easier heat exchange inside to outside than from zone to zone in the building.

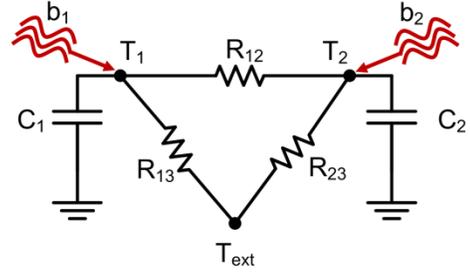

Fig. 4 Two-zone RC network plus heaters $b_1, b_2$.

TABLE 1
2-ZONE MODEL PARAMETERS

| Variable | Quantity | Numeric Value |
|---|---|---|
| $T_1(0)$ | Zone 1 initial temperature | 70 |
| $T_2(0)$ | Zone 2 initial temperature | 70 |
| $T_{ext}(0)$ | External zone initial temperature | 20 |
| $C_1$ | Zone 1 capacitance | 17 |
| $C_2$ | Zone 2 capacitance | 10 |
| $R_{12}$ | Zone 1/2 resistance | 150 |
| $R_{13}$ | Zone 1/3 resistance | 60 |
| $R_{23}$ | Zone 2/3 resistance | 100 |
| $b_1$ | Zone 1 heater output | 0.18 |
| $b_2$ | Zone 2 heater output | 0.22 |

Weather was formulated as the sum of two sinusoids plus an occasional temperature bias and additive random noise. The main sinusoid, 24 hour period and 20 degree peak-to-peak amplitude, provided daily fluctuation while the secondary sinusoid, 4 hour period and 5 degree peak-to-peak amplitude, provided slight variations throughout the day. Occasional temperature biasing mimicked hot-front and cold-front temperature swings. In order to demonstrate that the predictive controller was robust to non-perfect weather forecasts, MPC weather predictions were made from the daily sinusoid and bias but not the 4 hour sinusoid or the additive random noise.

### B. Unobservable Parameters

To better illustrate the reliability issues mentioned in the introduction and motivate the self-excitation and monitoring routines, examples of two failed parameter estimation attempts from particularly bad data sets were included (Fig. 5). Both examples were conducted by learning four RC parameters of a 2-zone model over a 3 day period: passive dynamics for the first day and active heating for remaining two days. Fig. 5.A, based on model in Table 1, shows one parameter whose covariance shrunk when heating commenced but whose estimate did not converge on the correct value. Fig. 5.B, based on a different model with small thermal resistance between zone 1 and 2, shows a parameter whose estimate became negative when heating commenced. These poor estimates are not an artifact of initial parameter seeding: notice how initializing the parameter value to be larger than the correct value did not prevent the estimator from guessing a negative value. Correcting such problems requires a better excitation signal in the measured data. In the proposed framework, the monitoring routine would detect a negative parameter or run a bank of filters to detect a lack of consensus on a parameter estimate and with self-excitation attempt to converge on a better parameter estimate.

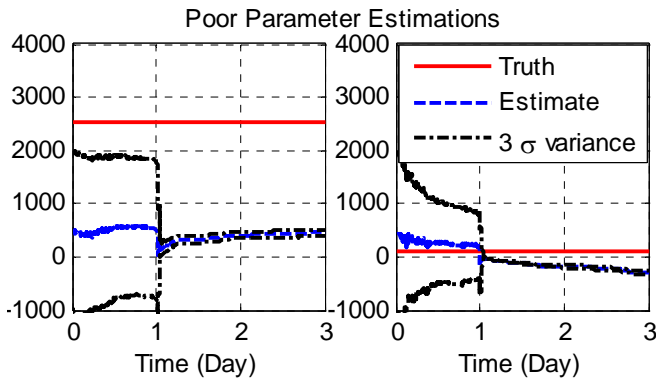

Fig. 5 Initial passive learning, heating commences at end of day one,
A.) *(left)* Parameter covariance shrinks with poor estimate,
B.) *(right)* Parameter estimate becomes negative.

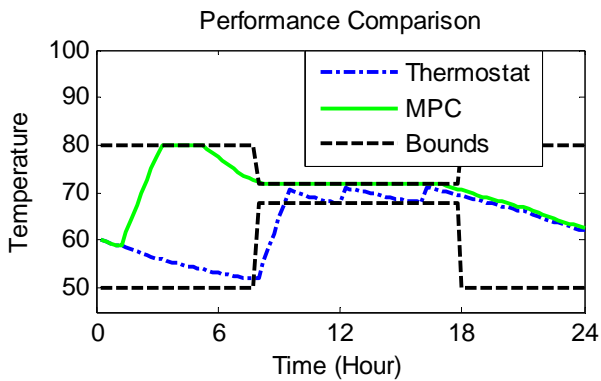

Fig. 6. MPC errant behavior is due to bad model parameters.

Thermal models with very bad parameter estimates cannot be reliably used for MPC. Using the parameters poorly estimated for the model in Table 1, a simulation was run comparing MPC performance against a baseline thermostat. MPC controlled zone 1 well but control of zone 2 was errant, see Fig. 6. The model contained bad parameters for the edge connecting zone 1 and zone 2, which caused MPC to attempt to apply extra heat to zone 1 by heating zone 2 to its upper temperature bound. MPC's poor performance, despite obeying temperature bounds, caused 30% more energy to be used compared to the well-tuned thermostat controller. Results such as these necessitate both self-excitation and monitoring routines to be part of the online estimation and control framework.

## V. SELF-EXCITATION AND MONITORING

### A. Overall Framework

Proper operation of MPC is dependent upon a good thermal model of the building. Due to the inherent low level of excitation typically present in most buildings, active learning is an essential component to a robust online system identification routine [4]. For actively heated and cooled buildings, this active learning can be accomplished by self-exciting the building's thermal dynamics. The goal is to find which system input to excite in order to reduce the uncertainty of a parameter. This can be accomplished by finding which zone is associated with a parameter of interest and exciting the zone's actuator. A combined learning, control, and self-excitation framework has been proposed in Fig. 3, which was built from a typical MPC system and augmented with the self-excitation components: Experiment Generator and Selector.

Functionally speaking the Generator provides a listing of parameters that are poorly known in the thermal model and the Selector chooses an appropriate time to actively learn those parameters by exciting the system. *Specifically the Experiment Generator uses covariance and model sensitivity criteria to correlate poorly known parameters, back to physical zones that can be excited. Given the building state and forecasted conditions, the Experiment Selector evaluates whether it makes sense to excite any physical zones, at a given time, by modifying temperature set points over a short horizon.* The goal is to not significantly disrupt occupants or waste energy while trying to improve the learned model. The following sections cover these active learning functions.

### B. Experiment Generator

The UKF estimates parameters based on measurements of the building temperatures and updates a covariance matrix for the set of parameters. The covariances can be thought of as an inverse information criterion for the estimated parameter. So a small covariance for a given parameter implies that parameter is known well. In general we assume the UKF covariances accurately represent the quality of the parameter estimate, that is, we assume the estimator is unbiased.

Recall that estimated parameters correlate with actual resistances and capacitances of physical zones in the building. *The Experiment Generator correlates poorly known parameters back to physical zones that can be excited.* We propose three methods for identifying zones to excite:
1. Covariance Eigen-decomposition
2. Variational temperature methods
3. Monte Carlo energy methods.

Below is an exploration of each method.

The covariance Eigen-decomposition method attempts to pick the most uncertain parameter subspaces and then make an ordered list of the zones which correlate to those parameters. Exciting the listed zones should provide information to improve the worst parameter estimates shrinking the largest covariances. By computing eigenvalues $\lambda$ and eigenvectors $v$ of the subset of the covariance matrix $P_{k|k}$ that is associated with RC parameters, the Eigen-decomposition creates in (8) a prioritized listing of parameter vectors that the estimator needs help to actively learn through excitation. The listing is ordered by largest singular-value which correlates with the most uncertain parameter set[2].

$$[\lambda, v] = eig(P_{RC\ k|k}) \quad (8)$$

To determine the appropriate excitation this listing of uncertain parameters must be correlated back to physical thermal zones as one cannot directly excite parameters. Thus the eigenvector $v_m$ for each eigenvalue $\lambda_m$ must be related back to the respective element of $A$ based on how $A$ was constructed in (3). Reciprocals of estimated RC parameters were used to build the $A$ matrix, but a parameter sensitivity analysis of this matrix would show that the primary factor of interest is solely which nodes have connected edges with large variances. Using (3) a simple correlation function (9) has been created to define temperature node uncertainty weight vector $N_m$ for each eigenvalue $\lambda_m$. External nodes have infinite capacitance, so no parameters are estimated for these nodes. This causes the row in $A_{\lambda_m}$ corresponding to any external node to have all zero elements. In order to represent excitation relative to an external node $i$, the $i$th column is summed and stored in the $(i,i)$ term of $A_{\lambda_m}$. The vector $N_m$ contains large values for nodes whose edge-connected parameters are poorly known. Here index $k$ correlates with parameter $C_iR_{ij}$ as defined by element $p_k$ in the vector $\bar{p}$ minimal parameterization. For example, in the simulation results parameter $p_1$ corresponds to $R_{12}C_1$, $p_2$ to $R_{13}C_1$, $p_3$ to $R_{12}C_2$ and $p_4$ to $R_{23}C_2$. If any redundant parameters were multiplicatively cancelled out in the minimal parameterization step, then their multiplication relation will need to be substituted back into the appropriate edge in (9) by performing the same multiplication on the Eigen-vector components.

$$A_{\lambda_m} = \begin{cases} A_{ij} = 0 & \text{if } i \neq j, (i,j) \notin E \\ A_{ij} = v_{m,k} & \text{if } i \neq j, (i,j) \in E \\ A_{ij} = \sum_{l \neq i} A_{il} - \sum_{l \neq i} A_{li} & \text{if } i = j \end{cases} \quad (9)$$

$$N_m = \text{diag}(A_{\lambda_m})$$

By analyzing vector $N_m$ we can see which nodes should be excited relative to each other. This information, the set of vectors $N$ and eigenvalues $\lambda$, is passed to the Experiment Selector which can actually create excitation by modifying temperature set points accordingly to create a difference in the temperature between nodes.

If the RC model was heavily biased, the covariance for the estimated parameter may be an unreliable metric of estimate quality. Additionally, all parameters may not significantly effect the temperature estimates of all zones in the building. Thus in certain situations the priority in excitation might be better posed by examining the sensitivity of zone temperatures to individual parameters. We call this a variational method which can be mathematically posed as a partial derivative of temperature with respect to parameters as the Jacobian of the state space system

---
[2] Singular values for a covariance matrix equals the absolute value of the eigenvalues.

$$\frac{\partial A}{\partial \bar{p}} = \begin{bmatrix} \frac{\partial T_1}{\partial p_1} & \cdots & \frac{\partial T_1}{\partial p_k} \\ \vdots & \ddots & \vdots \\ \frac{\partial T_i}{\partial p_1} & \cdots & \frac{\partial T_i}{\partial p_k} \end{bmatrix} \quad (10)$$

Parameters are also being estimated for the HVAC components, and the controller is using both the thermal dynamics and actuator constraints when selecting a control action. The energy consumption across zones may be imbalanced, so selecting the parameter that affects the most temperatures may not result in initially saving the most energy. Monte Carlo methods could be used to examine the sensitivity of energy consumption from the controller to parameter variations in the thermal model. By sampling parameter values, a Monte Carlo simulation could be run to examine $\partial E/\partial \bar{p}$, the partial derivative of energy with respect to the parameter values.

Whichever parameter's variance causes the largest changes in energy are the parameters that should be estimated best and warrant excitation. The main limitation to the variational and Monte Carlo methods is that they are being run on the simulated model. Given a bad initial model this method will not give meaningful insight. We recommend the Eigen-decomposition method for buildings with no strong prior model (e.g. any building where one expects to use a gray-box online learning method to acquire an initial model.) We recommend the variational method or Monte Carlo method for buildings which have a strong model, either after initial model convergence or from a strong Bayesian prior. These methods are applicable when one wishes to continue to refine their parameter estimates over time to capture the building's change in use, wear/aging, fault detection, or general purpose monitoring.

*C. Experiment Selector*

*The Experiment Selector evaluates the query: at the given time does it make sense to modify any temperature set points over a short horizon $h_s$ given current building state, occupancy, upcoming temperature bounds, weather predictions, and current thermal dynamics model?* During initial acquisition without a Bayesian prior, one main limitation is that the large uncertainty of the current estimated model may cause the dynamics and baseline control calculations to be erroneous. If the UKF has not converged on parameter estimates then the MPC control will be unreliable and any routine that uses the estimates is liable to produce erroneous behavior—thus, until parameters have converged, thermostat control should be used. For new buildings an initial model is often available from the designers to seed the UKF. For retrofit buildings data may be collected until model convergence before running MPC. In any case, properly introducing excitation into the system reduces this learning time.

In software simulation it is common to see pseudo-random binary inputs used for excitation to generate control-oriented models from high fidelity plant simulations. While this works great in simulation, pseudo-random binary excitation signals subject chiller, pump, or fan hardware to unnecessary wear and tear. One could create bounded pseudo-random binary inputs for specific components that have been examined to ensure they can handle the input spectrum [21]. The cost of energy savings over a month or even year timeframe is often significantly less than the replacement cost for one broken component. That is, the risk is too high to use wide bandpass excitation signals, therefore we seek strategically targeted excitation signals. Keeping this in mind we present two experiments, one simple heuristic for perturbing temperature values and a model predictive option.

The heuristic options looks at the first vector $N_m$ and picks out the two largest values—let $i$ and $j$ contain their respective node ids. If the values are within 25% of each other, then: case 1) excitation is desired between node $i$ and $j$; otherwise: case 2) excitation of node $i$ relative to all other nodes is desired. Given the current temperatures, predicted weather, and temperature bounds, determine if heating (or cooling) the desired nodes will cause a noticeable increase in the temperature difference of interest. If so, then apply heat for four subsequent time steps or until an upper (or lower) temperature bound is reached.

After initial acquisition, when MPC is running, the decision and temperature modification can be formalized as a convex optimization problem using the format developed for MPC. To seed the problem, results from the current MPC solver iteration provides baseline temperature differences and baseline control effort for the original temperature bounds. The optimization problem is configured to maximize the temperature difference of interest by varying individual zone temperature bounds, while observing thermal dynamics, actuation saturation limits, hard constraints on the original temperature bounds, and a threshold (e.g. 110%) control effort compared to the unexcited scenario. Mathematically this can be written as (11). Note that the short horizon $h_s$ over which temperature bounds $e$ are being modified should be several times smaller than the control horizon $h$ in order to have a fair basis of comparison for expected control effort under excitation—the excitation may have residual effect on dynamics that is predictable if $h \gg h_s$ and $length(e) = h_s$.

*Maximize J:*

$$J = \frac{1}{n}\sum_{k=1}^{h}|T_i(k) - T_j(k)|$$

where $i$ and $j$ are nodes of interest
*Subject to:*

$$\begin{aligned}
T(k+1) &= AT(k) + B_{ext}T_{ext}(k) + B_{ctrl}u(k)\\
0 &\leq u_i(k) \leq 1\\
T(k) &\geq r_{\min}(k)\\
T(k) &\leq r_{max}(k)\\
\sum_{k=0}^{h-1} u(k) &\leq 1.1 * u_{baseline}\\
e(k) &\geq r_{\min}(k)\\
e(k) &\leq r_{max}(k) - 2\\
T(k) &\geq e(k)
\end{aligned} \quad (11)$$

*Given numerical values for vectors:*

$$T_{ext}, T(0), r_{max}, r_{min}, r, u_{baseline}, i, j$$

If the excited temperature difference $J$ minus the baseline temperature difference is above a threshold then the experiment should be run. This selection routine is then repeated for each smaller eigenvalue. If no selection is made then the threshold is slightly decreased for the next timestep iteration in order to promote occasional excitation experiments. When a selection occurs, then the excitation is run until the horizon $h_s$ expires and the threshold is reset.

### D. Monitoring

We showed in [3] that parameter estimates were robust to solar radiation and ground coupling disturbances, however, while performing this paper's study, heating loads were observed to occasionally degrade the stability of RC parameter estimates. Two problems occasionally arose: physics violations and false convergence. Physics violations occur when an RC parameter is estimated to be negative violating conservation of energy. False convergence occurs when the covariance errantly shrinks while the filter is doing a poor job of estimating a parameter. Shrunken covariance can prevent future data from correcting a parameter estimate. Detecting the former is easier than the latter, but both can be ameliorated with an auxiliary monitoring routine. In this study a human operator monitored performance.

While an estimator is running the monitoring routine occasionally saves the current parameter estimates and covariances. By simply stopping estimation when a parameter gets close to zero or becomes negative the estimator can be re-initialized to the last saved checkpoint. This allows poor data sections to be thwarted from corrupting estimates.

False convergence can be combatted with consensus testing. Every so often a new copy of the UKF can be initialized with different initial parameter seeds and run in parallel with the original UKF filter(s). Parameter values are compared based on their expected value and covariance—how many standard deviations away are the other filter's estimates of that parameter. Typically a consensus filter can increase the robustness of initial acquisition routines to poor initial data and poor initial seed estimates. Correcting these numerical issues was outside the scope of this study.

## VI. MATLAB SIMULATED BUILDING RESULTS

### A. Excitation and MPC Performance

Running the same 3 day simulation on the model from Table 1, with the addition of excitation, significantly improved parameter estimates. Fig. 7 shows the improved estimation of one RC parameter. Notice when excitation was applied during the third day, the parameter estimate is improved to match the true value and the covariance shrinks.

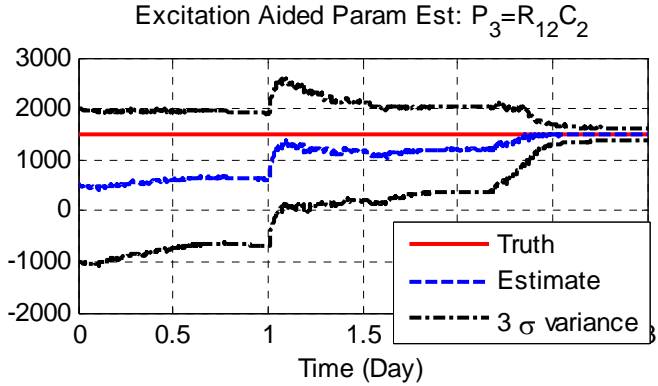

Fig. 7. Significant estimation improvement due to excitation. Day 1 passive thermal response, day 2 uniformly heated building, day 3 excitation applied.

A weeklong control simulation comparison was run between the well-tuned thermostat controller and MPC using the model acquired after excitation. Weather was varied throughout the week from averaging in the mid-fifties Fahrenheit for the first couple days to the teens for the last couple days. Table 2 contains the numeric results of the simulation. The MPC used 7.5% less energy than the thermostat controller while improving comfort, defined as the temperature error term from (7), by almost 50%. In this 2-zone system it was easy to observe in Fig. 8 where MPC outperformed the thermostat: on cold days the thermostat would turn on too late, on warm days the thermostat would turn on too early.

### B. Analysis of Excitation Effects

An observability analysis was performed to provide insight into the mechanics of the estimation algorithm's performance during excitation. The true RC parameters for the system were known and constant but the temperature states varied through time. For the 2-zone house there were 3 temperatures, 4 unique RC parameters, and 2 disturbance parameters for the heaters. Because disturbances were estimated accurately and only learned while the heating system was on, they were omitted from this analysis. This study focuses on the observability of the RC building dynamics.

In order to build an observability matrix—a linear approximation of the system parameter's observability, discrete-time $F$ and $C$ matrices were formed. The state space $F$ matrix is built as a state transition matrix by calculating the matrix exponential of the Jacobian of the dynamics for the augmented state and parameters vector.

$$F(k) = \Phi(k+1, k) = e^{\left(\frac{dA}{dx}\big|_{x(k)}\right)}$$

$$\frac{dA}{dx} = \begin{bmatrix} \frac{df_1}{dT_1} & \cdots & \frac{df_1}{dp_4} \\ \vdots & \ddots & \vdots \\ \frac{df_7}{dT_1} & \cdots & \frac{df_7}{dp_4} \end{bmatrix} \quad (12)$$

The state space $C$ matrix is simply an identity matrix for the measured temperature states padded with zeros.

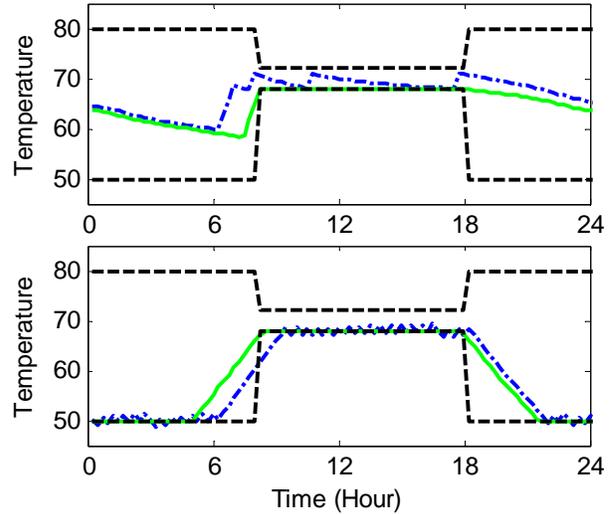

Fig. 8. A.) (top) Warm external temp, thermostat heating too early, B.) (bottom) Cold external temp, thermostat heating too late. See legend from Fig. 6.

TABLE 2
MPC VERSUS THERMOSTAT CONTROL

| Controller | Statistic | Result |
|---|---|---|
| Thermostat | Occupant Discomfort | 0.170 |
| Thermostat | Energy Usage | 415 |
| MPC | Occupant Discomfort | 0.086 |
| MPC | Energy Usage | 384 |

$$C = [I_{3\times3} \quad 0_{3\times4}] \quad (13)$$

A discrete time observability matrix was built using the standard form $\mathcal{O}(k) = [C \quad CF(k) \quad \cdots \quad CF(k)^6]$. For the entire simulation the matrix had a maximum rank of 5 meaning 2 states were unobservable in any individual temperature configuration. The condition number varied in the range from $10^{16}$ to $10^{21}$ but gave little additional insight. A plot of the reciprocal of the condition number is shown in Fig. 9. However, by looking more closely at the components of the two nullspace vectors $v_1$, $v_2$ of the observability matrix, we can observe that as the excitation caused zone temperatures to vary, the primary axis directions of the nullspace changed as can be seen in Fig. 10. Specifically we observe that the excitation applied on the third day caused a significant increase in information about parameters $p_1 = R_{12}C_1$ and $p_3 = R_{23}C_2$ and explains the estimate improvement for those parameters.

The excitation did not improve the *number* of parameters that can be simultaneously estimated at one instant, but rather changed *which* parameters we could estimate at a given instant. Although the entire building may not be fully observable at any individual time step, over time, by retaining prior learned information, the entire building's model may be inferred. In other application areas, such as magnetometer based spacecraft attitude estimation, the system's dynamical evolution coupled with a Bayesian estimator are often used to fully observe systems that were initially thought to partially observable [31]. In summary the building system is not time-invariant observable, but it is time-varying observable given appropriate system excitation using the described Experiment Generator and Selector.

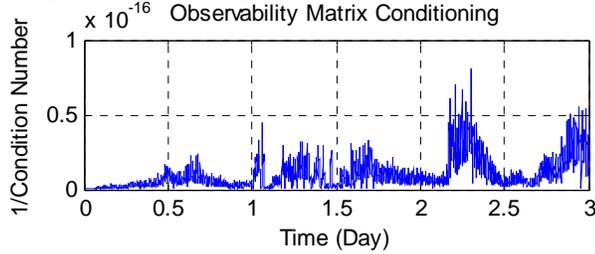

Fig. 9 Observability matrix conditioning over time.

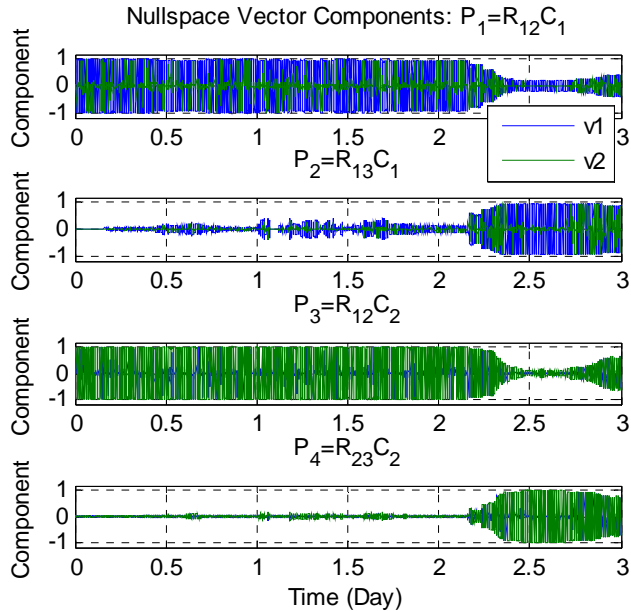

Fig. 10 Nullspace component magnitudes over time.

## VII. DISCUSSION

The main contribution of this paper was the presentation of methods for self-excitation of the HVAC system of a building. A simple thermostat controlled 2-zone building needed excitation in order to learn a gray-box thermal model given no prior parameter estimates. One of the more insightful observations was finding that excitation did not increase the total number of instantaneously observable parameters in the system but rather changed which specific parameters inference could be drawn on from the data. In practice the amount of natural excitation from external disturbances and the HVAC system will vary across buildings. Given that the simple 2-zone model presented here required excitation, was not time-invariant observable, and in agreeance with a number of studies done with real-data, we believe most buildings will require some form of excitation to learn control-oriented models. More complex buildings are likely to be in even greater need of excitation for proper identification. The specific implementation details of the thermal model estimation or predictive controller could be modified and refined, but there will always be a need for useful excitation and robust monitoring if such a system will ever be deployed across real buildings.

The Experiment Generator and Experiment Selector methods presented in this paper intuitively capture the general intent of self-excitation for buildings and worked for initial acquisition of the 2-zone building presented here. The next steps would be to evaluate their performance and computational feasibility in a co-simulation study where MLE+ links Energy Plus to Matlab. Most co-simulation model-inference studies assume infinite hardware response from the simulation and use pseudo-random binary inputs. Running the self-excitation in software while obeying real physical hardware constraints would enable tuning of the various self-excitation routines. Lastly this self-excitation routine will need to be tested on real hardware in real buildings.

The simulation conducted in this study demonstrated the robustness MPC building control has to weather uncertainty, while utilizing a dynamical model learned from an online gray-box estimation method. MPC demonstrated better occupant comfort and reduced energy use versus thermostat control.

## VIII. CONCLUSION

For a 2-zone building a technician may be able to tune the thermostat for a couple common scenarios but in more complex buildings with more zones, disturbances, and constraints, operator tuning becomes infeasible. Our simulation results, where MPC used an online learned model to outperform a well-tuned thermostat control, demonstrate the potential that online autonomous methods have in building systems. Various online methods were presented for excitation, and the effects of excitation on the estimation filter were analyzed. By coupling gray-box estimation and predictive control algorithms with online self-excitation and monitoring into a single framework, we have highlighted the potential for low-cost scalable methods to save energy and improve occupant comfort in buildings.